\documentclass[12pt,preprint]{aastex}

\shorttitle{Erupting Classical Nova in Globular Cluster}
\shortauthors{Shara et al.}

\begin{document}

\def\putplot#1#2#3#4#5#6#7{\begin{centering} \leavevmode
\vbox to#2{\rule{0pt}{#2}}
\includegraphics{#1}

\end{centering}}


\title{An Erupting Classical Nova in a Globular Cluster of M87}


\author{Michael M. Shara and David R. Zurek}
\affil{Department of Astrophysics, American Museum of Natural History, 
New York, NY 10024}

\author{Edward A. Baltz}
\affil{Astrophysics Laboratory, Columbia University, New York, NY 10027}

\author{Tod R. Lauer}
\affil{National Optical Astronomy Observatories, Tucson, AZ 85726}

\author{Joseph Silk}
\affil{University of Oxford, Oxford, United Kingdom 0X1 3RH}



\begin{abstract}
Only one certain classical nova eruption has ever been detected inside a
globular cluster - nova 1860 A.D. (T Sco) in M80. During a survey of
M87 we have detected an erupting star coincident (to within 0.08 pixels)
with a globular cluster of that giant elliptical galaxy. We are able to
discount variables in the foreground or background of M87. The light curve
and color of the erupting star match those expected for a nova at the 
distance
of M87. The chance superposition of an M87 field nova on the globular 
cluster
is very unlikely but cannot be completely ruled out.
Our detection hints at a globular cluster nova frequency $f \sim .004 $ 
novae/cluster/year,
much higher than previous observations have suggested.  

\end{abstract}



\section{Introduction}

The realization that X-ray sources are a thousand times more common in
globular clusters than in the rest of the Galaxy (\citet{cla75})
first spurred theorists to consider the effects of very close stellar 
encounters.
Tidal captures may occur during very near-misses of degenerate stars 
(\citet{fpr75}) and low mass main sequence stars. Neutron 
stars
with captured companions give rise to the observed cluster X-ray sources,
but white dwarfs are much more common than neutron stars. Thus dozens of 
tidal-capture
white dwarf - red dwarf binaries (cataclysmic variables, including classical
novae = CN) are also expected in dense globular cluster cores 
(\citet{rap94}).
Nearly three decades of predictions (e.g. \citet{fpr75}, \citet{kro84})
and surveys (e.g. \citet{gri01}, \citet{kni02}) have finally begun to reveal large numbers of cataclysmic
binaries (CVs) concentrated in the cores of Galactic globular clusters. 
Contrasting
tidal capture-generated novae with binary evolution-generated novae would be
very interesting. For example, do the two populations of novae outburst
with similar frequencies? It is clearly necessary to first find at least 
a few
tidal capture novae to begin such a study.

Over the past century, of order a thousand erupting CN have been detected in
the Galaxy and M31. However, only two candidate CN eruptions have
ever been observed in Galactic globular clusters (GC) (\cite{lut60} and 
\cite{hw64}).
Of these only nova 1860 (T Sco in M 80) seems beyond any doubt to be a 
cluster
member. This is based on the remarkable and independant discoveries and
light curve documentation by \citet{au62} and by \citet{pog60}, 
supplemented
by \citet{saw38} and \citet{wsb90}. \citet{sd95}
have recovered the quiescent nova in HST images of the core of M 80.
The second candidate (nova 1938 in M 14) was a genuine variable that had
the right brightness to be a CN, but details of its light curve are far
too scant to determine its true nature (\citet{hw64}). Recovery attempts 
have
been unsuccessful (\citet{sha86}; \citet{mar91}).

If classical novae were as overabundant in globular clusters
(relative to the field) as are X-ray sources we should have detected about
100 CN in the GC of the Galaxy and M31 over the past century. While 
there might
be one or more CN depletion mechanisms in GC, the present study
suggests instead that searches for CN in GC are very incomplete.

In the course of an HST imaging survey of M87 effectively spanning $\sim90$ 
days we have recently detected an
erupting CN coincident with a GC of that galaxy. 
If the nova rate in M87 GC is similar to that in the Galaxy we would expect to 
observe a nova only once in $\sim 20$ years of HST observations (see section 4).
The rarity of detected Galactic and M31 GC CN,
the ease with which the M87 nova was detected, and the possible 
implication for
globular cluster nova rates prompt us to report it here; full details 
and further
searches will be described elsewhere.  

\section{Observations \& Astrometry}

M87 was imaged by HST on the $30$ successive days from 28 May 2001 through
25 June 2001 inclusive. A log of the observations, taken through the 
F814W and
F606W filters, is given in Table 1. We have also examined all the HST 
archival
images of M87 and its globular clusters, so that our effective survey 
time is about 90 days (see below). Most other HST images of M87 
and its globular clusters
are single-epoch. Erupting classical novae brighten from M $\sim +4$ to 
M $\sim -8$ in about one day. We would have thus detected the nova 
reported here in a 
similar, average luminosity M87 GC at maximum or within the $\sim5$ days 
following maximum light.
Thus the effective survey time for the 10 other epochs is $\sim6$ days each, 
and for the entire survey it is $\sim30+60 = 90$ days.
The M87 GC in which we detected
a variable is marked in the finder chart shown in Figure 1. The GC is
located at $\alpha_{(2000)}$=12:30:45.4, $\delta_{(2000)}$=12:23:16.1 as
determined using the world coordinate system in the WFPC2 header. The GC is
located 13.51 arcseconds south and 60.95 arcseconds west of the M87 core.
While the GC is quite small and faint, we have determined that it is
a resolved object and not a star. The globular cluster appears at $I = 23.06$,
corresponding to $M_I = -8.06$ ... near the mean luminosity of globular 
clusters.
The cluster displays $V - I = +1.0$ ,  also typical of globular clusters.

In Figure 2 we show a series of five images centered on the position noted
above. We deliberately added, at the left of Figure 2, an image taken 
six years
prior to the others. This (and other epoch images) clearly shows the 
presence
of the M87 GC {\it without} the variable. The circled GC, coincident 
with the
variable, is close to, and significantly fainter than another GC above and to
the right in Figure 2. The middle image of Figure 2 (8 June 2001) shows the
variable at maximum light: it has effectively brightened its host GC to 
match
the nearby GC. Towards the end of the 30 day observing campaign the 
variable
has faded from view and the circled GC is again faint compared to its 
neighbor.

The light curve of the GC plus variable is shown in Figure 3. The hint of a
brightness increase on day 11 is strongly confirmed by the brightness peak
seen in both F814W and F606W on day 12. The GC remains obviously brighter
than its quiesent state for $\sim 6$ days after maximum light in the F606W
filter and $\sim 3$ days in the F814W images. The photometry of Figure 3 is
listed in Table 1.  

\section{Is the Variable a Nova?}

Have we discovered an erupting nova in an M87 GC? We show below that the
light curve is fully consistent with a CN in M87. However, to answer the
question in the affirmative we must first rule out other possibilities. 
Most
likely amongst these are: a chance superposition of a background 
supernova, a
gamma ray burst, an M87 nova, a microlensing event or a Milky Way variable
along the line of sight to the GC. The supernova hypothesis can immediately
be discarded because of the rapid rise and decline times of the light curve.

The detailed shape of the decline from maximum light on day 12 through day
21 suggests a power law decline L $\alpha$ $t^{-1.0}$. There are observed
optical afterglows of gamma ray bursts (GRB) with similar decline laws
(e.g. GRB000301C \citet{jen01} and GRB980703 \citet{hol01}). GRB afterglows
typically achieve R or I $\sim 20$ one
day after outburst; the variable in M87 is 3 magnitudes fainter. One might
thus argue that the outbursting object is at higher redshift than most
GRB afterglows observed so far. We regard this explanation as very unlikely
both because of the rarity of GRB ($\sim 1$/day over the entire sky) and
because there is some indication of a precursor rise in the F606W image on
day $11$.

The constant brightness of the GC in images previous and subsequent to the
eruption, the rapid brightness rise and several day decline, and the
brightness and moderately blue color of the variable rule out most
types of Galactic variables. RR Lyraes, eclipsing binaries and flare stars
don't match the observed brightness profile and/or blue color.
If the variable is a dwarf nova along the line of sight to M87 it is about
100 kpc distant from the Galaxy...an intergalactic tramp. We regard this  
possibility as extremely unlikely. The asymmetry of the light curve likely
rules out a microlensing event.

In F606W the brightening of the GC is $\sim 1.2$ magnitudes. Taking into
account the brightness of the GC the peak mag of the variable in F606W is
$23.57$. At the distance of M87 $(m-M) \sim 31.12$ (\citet{kun99})
this corresponds to an absolute magnitude of $-7.55$. The slope of
the variable's decline yields decline times of $t_2 \sim 18$ days
and $t_3 \sim 27$ days. Using the absolute mag - decline time relations
for novae from \citet{dd00} yields $M_V \sim -8.35$ ($t_3$ relation) and
$M_V \sim -8.12$ ($t_2$ relation), in good agreement with the observed 
value. The moderately blue color of the variable (at peak brightness it 
displays V - I = 0.61) further supports the suggestion of a classical nova.
 
The most difficult possibility to rule out is the chance superposition of
an M87 {\it field} nova on the GC. We have been quite successful in
detecting erupting novae in M87 (\citet{sz02a} and \citet{sz02b}).
About 3 to 5 CN in eruption ($\sim1$ or $2$ on each WF chip) are 
detectable with HST in F814W band images
every time one looks. Novae are usually much brighter in ultraviolet or blue
passbands so the number in eruption in each epoch may be considerably higher. 
This allows us to crudely estimate the probability of
a chance GC-CN superposition. Each WF chip has $562,500$ useful
pixels, and about $0.5\%$ are covered by M87 GCs. Randomly distributing 
1 nova on a WF chip leads to a chance cluster-nova miss rate of $0.995$ 
and thus a chance coincidence rate of $1 - 0.995 = 0.005$. The near perfect 
positional coincidence of the GC and the nova candidate 
(see Figure 4; $\pm 0.08$ pixels, corresponding to 0.6 pc at the distance 
of M87) make the chance coincidence rate $\sim 100$ times better ... i.e. 
about $0.0005\%$. We understand the dangers of {\it a posteriori} statistics, 
and recognize that we may well have detected a fluke superposition of a 
field CN on an M87 GC. In the rest of this letter, though, we will assume 
that the CN is a genuine globular cluster member.

\section{Nova Rates}

The detection of one certain erupting CN in the $\sim 150$ GC of the
Milky Way over the past $\sim 140$ years places a rough lower limit on
the nova-in-cluster eruption frequency f: $f > .00005$ CN/cluster/year.
Weekly observations are required to find the fastest novae, but most 
Galactic globular
clusters have not been observed this frequently (by either professional or
amateur astronomers). Erupting CN rival their host globular clusters in
luminosity and should draw attention to themselves, but only to amateur and
professional astronomers realizing that a very bright star in a globular 
cluster
is unusual. The incompleteness factor in f is difficult to estimate, but
it must be at least a factor of 2 (due to seasonal and weather effects)
and is probably much more.

Motivated by theory and models suggesting the existence of many
cataclysmic variables (CVs) in GCs, \citet{ctp90} and \citet{tcs92} surveyed
large numbers of M31 globulars for erupting CN. None were detected in an 
H$\alpha$
survey over an effective survey time of one year for the entire known 
M31 GC
system. This places a useful upper limit on f: $f < .005$ CN/cluster/year.
(We note in passing that while the range of observed globular cluster 
luminosities
and masses is substantial, the overall Galaxy and M31 globular cluster 
systems are
remarkably similar (\citet{rac79}).

The survey we are reporting here had an effective time coverage of 90 days
for the 1057 globular clusters of M87 reported by \citet{kun99}. This
one detected erupting nova in our survey yields a GC nova frequency 
f$\sim .004$ CN/cluster/year,
just slightly smaller than the upper limit reported by \citet{tcs92}.
Unless we were very lucky to have detected a GC CN
(or the CN is not a GC member) this suggests that
classical nova eruptions in globular clusters are up to 100 times
more common than current detections in the Milky Way suggest. A search
in M31 extended in time by a factor of about 3-5 over that of \citet(tcs92) 
should reveal two or more GC novae, and firm up or refute the tentative
(and surprisingly high) rate we report in this letter. While a 
twice-weekly search of every globular cluster in the Milky Way is a daunting 
task, it is not beyond the capabilities
of modest robotic telescopes.  A rate of .004 novae/yr/cluster suggests 
a remarkable overall nova rate of 1 event every 2 years in the globulars 
of the Milky Way...well worth searching for.

\acknowledgments

We gratefully acknowledge support under grant GO 8592 from the Space 
Telescope
Science Institute. This research is based on observations made with the
NASA/ESA Hubble Space Telescope obtained at the Space Telescope Science 
Institute.
STScI is operated by the Association of Universities for Research in 
Astronomy, Inc.  under NASA contract NAS 5-26555.

\clearpage


\begin{figure}[ht]
\putplot{M87_fig1.ps}{4.0in}{0}{90}{90}{-260}{-270}
\putplot{image_patch3.ps}{3.0in}{0}{70}{70}{-60}{-100}
\caption{Finder chart of the erupting classical nova in M87 which is 
coincident with a globular cluster. The square in the lower left quadrant 
of the HST WFPC2 image is shown magnified in the overlay to its right, in
which the GC host is indicated with an arrow. See figure 2 for a closeup. \label{fig1}}
\end{figure}

\clearpage

\begin{figure}[ht]
\putplot{gcnova_epochs2.ps}{4.0in}{0}{100}{100}{-250}{-270}
\caption{Closeup of the erupting classical nova coincident with the
globular cluster of M87. Left image is from a F555W WFPC2 frame
taken 1995; the four rightmost images, taken in 2001, show the rise
and decline of the nova (circled with its host globular). A second,
brighter field globular cluster of M87 is above and to the right of the
circled nova and GC. \label{fig2}}
\end{figure}

\clearpage

\begin{figure}
\putplot{glob_lightcurve3.eps}{3.0in}{0}{70}{70}{-250}{0}
\caption{F606W and F814W light curves of the M87 nova and globular cluster
combined. Magnitudes are in the ST system. \label{fig3}}
\end{figure}

\clearpage

\begin{figure}
\putplot{comparison2.ps}{3.0in}{0}{100}{100}{-260}{-170}
\caption{Comparison of the globular cluster (center of each panel) without 
the nova erupting (left) and with the nova erupting (right). The distribution 
of light during eruption is consistent with the nova being superposed directly 
on the GC. These subimages are taken from the interlaced F814W frames; the pixel
scale is twice as fine as the F606W images shown in Figure 2. The left panel
is the average of all F814W images that do not have the nova in outburst. The
right panel is the F814W image from day 12. \label{fig4}}
\end{figure}






\clearpage

\begin{deluxetable}{ccccccc}
\tabletypesize{\scriptsize}
\tablecaption{Observations and Photometry (30 days beginning 2001/05/28 and ending 2001/06/25) \label{tbl-1}}
\tablewidth{0pt}
\tablehead{
\colhead{Day} & \colhead{}   & \colhead{F814W}   & \colhead{} &
\colhead{}  & \colhead{F606W} & \colhead{} \\
\colhead{} & \colhead{EXPTIME} & \colhead{ST Mag} & \colhead{Error} &
\colhead{EXPTIME} & \colhead{ST Mag} & \colhead{Error} 
}
\startdata
 1 & 4 X 260.0 sec &  24.47 &  0.11 & 1 X 400.0 sec &  24.27 &  0.11\\  
 2 & 4 X 260.0 sec &  24.41 &  0.11 & 1 X 400.0 sec &  24.24 &  0.11\\
 3 & 4 X 260.0 sec &  24.11 &  0.20 & 1 X 400.0 sec &  24.18 &  0.10\\
 4 & 4 X 260.0 sec &  24.37 &  0.13 & 1 X 400.0 sec &  24.38 &  0.12\\  
 5 & 4 X 260.0 sec &  24.26 &  0.07 & 1 X 400.0 sec &  24.35 &  0.12\\
 6 & 4 X 260.0 sec &  24.40 &  0.09 & 1 X 400.0 sec &  24.23 &  0.11\\
 7 & 4 X 260.0 sec &  24.13 &  0.16 & 1 X 400.0 sec &  24.09 &  0.09\\
 8 & 4 X 260.0 sec &  24.34 &  0.11 & 1 X 400.0 sec &  24.48 &  0.13\\
 9 & 4 X 260.0 sec &  24.35 &  0.15 & 1 X 400.0 sec &  24.35 &  0.12\\
10 & 4 X 260.0 sec &  24.38 &  0.18 & 1 X 400.0 sec &  24.24 &  0.11\\  
11 & 4 X 260.0 sec &  24.23 &  0.05 & 1 X 400.0 sec &  24.07 &  0.09\\
12 & 4 X 260.0 sec &  23.51 &  0.07 & 1 X 400.0 sec &  23.15 &  0.04\\
13 & 4 X 260.0 sec &  23.83 &  0.08 & 1 X 400.0 sec &  23.27 &  0.05\\
14 & 4 X 260.0 sec &  24.08 &  0.08 & 1 X 400.0 sec &  23.82 &  0.07\\
15 & 4 X 260.0 sec &  24.14 &  0.09 & 1 X 400.0 sec &  23.82 &  0.07\\
16 & 4 X 260.0 sec &  24.22 &  0.17 & 1 X 400.0 sec &  23.98 &  0.09\\
17 & 4 X 260.0 sec &  24.16 &  0.10 & 1 X 400.0 sec &  24.03 &  0.09\\
18 & 4 X 260.0 sec &  24.33 &  0.07 & 1 X 400.0 sec &  23.94 &  0.08\\
19 & 4 X 260.0 sec &  24.37 &  0.16 & 1 X 400.0 sec &  24.12 &  0.10\\
20 & 4 X 260.0 sec &  24.33 &  0.11 & 1 X 400.0 sec &  24.11 &  0.10\\
21 & 4 X 260.0 sec &  24.47 &  0.08 & 1 X 400.0 sec &  24.33 &  0.11\\
22 & 4 X 260.0 sec &  24.15 &  0.06 & 1 X 400.0 sec &  24.22 &  0.11\\
23 & 4 X 260.0 sec &  24.38 &  0.14 & 1 X 400.0 sec &  24.23 &  0.11\\
24 & 4 X 260.0 sec &  24.29 &  0.18 & 1 X 400.0 sec &  24.16 &  0.10\\
25 & 4 X 260.0 sec &  24.26 &  0.09 & 1 X 400.0 sec &  24.26 &  0.11\\
26 & 4 X 260.0 sec &  24.33 &  0.09 & 1 X 400.0 sec &  24.29 &  0.11\\
27 & 4 X 260.0 sec &  24.36 &  0.12 & 1 X 400.0 sec &  24.43 &  0.13\\
28 & 4 X 260.0 sec &  24.34 &  0.22 & 1 X 400.0 sec &  24.45 &  0.13\\
29 & 4 X 260.0 sec &  24.31 &  0.08 & 1 X 400.0 sec &  24.20 &  0.10\\
30 & 4 X 260.0 sec &  24.19 &  0.10 & 1 X 400.0 sec &  24.04 &  0.09\\
\enddata

%
%

\end{deluxetable}


\begin{thebibliography}{}

\bibitem[Auwers (1862)]{au62} Auwers, G.F. 1862, Astron.Nach., 58, 374
\bibitem[Ciardullo, Tamblyn \& Phillips (1990)]{ctp90} Ciardullo, R., Tamblyn, P. \& Phillips, A.C. 1990, \pasp, 102, 1113
\bibitem[Clark (1975)]{cla75} Clark, G. 1975, \apj, 199, L143
\bibitem[Downes \& Duerbeck (2000)]{dd00} Downes, R.A. \& Duerbeck, H.W.  2000, \aj, 120, 2007
\bibitem[Fabian, Pringle \& Rees (1975)]{fpr75} Fabian, A.C., Pringle, J.E., \& Rees, M.J. 1975, \mnras, 172, 15
\bibitem[Grindlay et al (2001)]{gri01} Grindlay, J.E., Heinke, C., Edmonds, P.D., \& Murray, S.S. 2001, Science, 292, 2290
\bibitem[Hogg \& Wehlau (1964)]{hw64} Hogg, H.B. \& Wehlau, A. 1964, JRASC, 58, 374
\bibitem[Holland et al (2001)]{hol01} Holland, S., Fynbo, J.U., Hjorth, J., et al 2001, \aap, 371, 52
\bibitem[Jensen et al (2001)]{jen01} Jensen, B.L., Fynbo, J.U., Gorosabel, J., et al 2001, \aap, 370, 909
\bibitem[Knigge et al (2002)]{kni02} Knigge, C., Zurek, D.R., Shara, M.M., \& Long, K.S. 2002, \apj, 579, 752
\bibitem[Krolik (1984)]{kro84} Krolik, J.H. 1984, \apj, 282, 452
\bibitem[Kundu et al (1999)]{kun99} Kundu, A., Whitmore, B.C., Sparks, W.B., Machetto, F.D., Zepf, S.E. \& Ashman, K.M. 1999, \apj, 513, 733
\bibitem[Luther (1860)]{lut60} Luther, E. 1860, Astron.Nach., 53, 293
\bibitem[Margon et al (1991)]{mar91} Margon, B., Anderson, S.F., Downes, R.A., Bohlin, R.C. \& Jacobsen, P. 1991, \apj, 369, L71
\bibitem[Pogson (1860)]{pog60} Pogson, N.R. 1860, \mnras, 21, 32
\bibitem[Racine \&Shara (1979)]{rac79} Racine, R. \&Shara, M. 1979, \aj, 84, 1694
\bibitem[Rappaport \&di Stefano (1994)]{rap94} Rappaport, S. \& di Stefano, R., 1994, \apj, 423, 274
\bibitem[Sawyer (1938)]{saw38} Sawyer, H.B. 1938, JRASC, 32, 69
\bibitem[Shara \& Drissen (1995)]{sd95} Shara, M.M., \& Drissen, L.  1995, \apj, 448, 203
\bibitem[Shara et al (1986)]{sha86} Shara, M.M., Potter, M., Moffat, A.F.J., Hogg, H.S. \& Wehlau, A. 1986, \apj, 311, 796
\bibitem[Shara \& Zurek (2002a)]{sz02a} Shara, M.M. \& Zurek, D.R. 2002a, The Physics of Cataclysmic Variables and Related Objects, ASP Conference Proceedings, Vol. 261. Eds. B. T. Gänsicke, K. Beuermann, and K. Reinsch.   
\bibitem[Shara \& Zurek (2002b)]{sz02b} Shara, M.M. \& Zurek, D.R. 2002b, in Classical Nova Explosions, AIP Conference Proceedings, Vol. 637.  Eds. Margarita Hernanz and Jordi José.  p.457-461
\bibitem[Tomaney, Crotts \& Shafter (1992)]{tcs92} Tomaney, A., Crotts, A. \& Shafter, A. 1992, AAS, 181, 7309
\bibitem[Wehlau, Sawyer-Hogg \& Butterworth (1990)]{wsb90} Wehlau, A., Sawyer-Hogg, H., \& Butterworth, S. 1990, \aj, 99, 1159


\end{thebibliography}
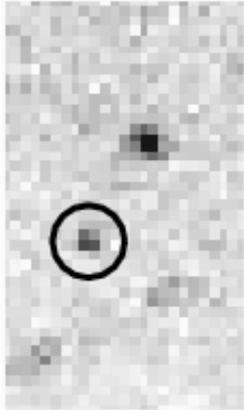
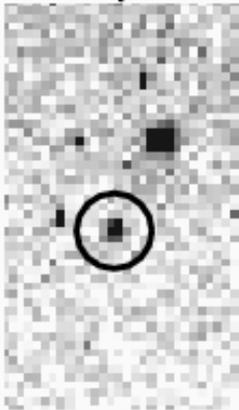
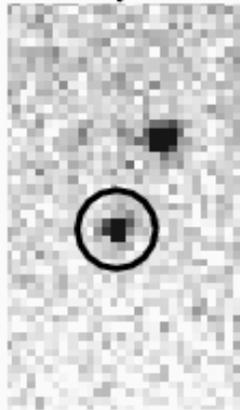
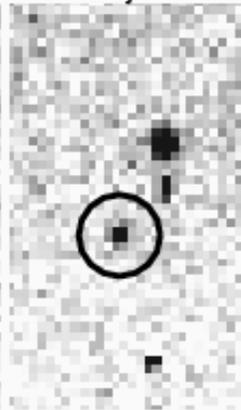
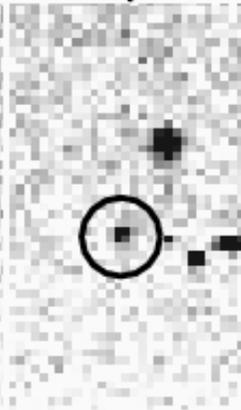
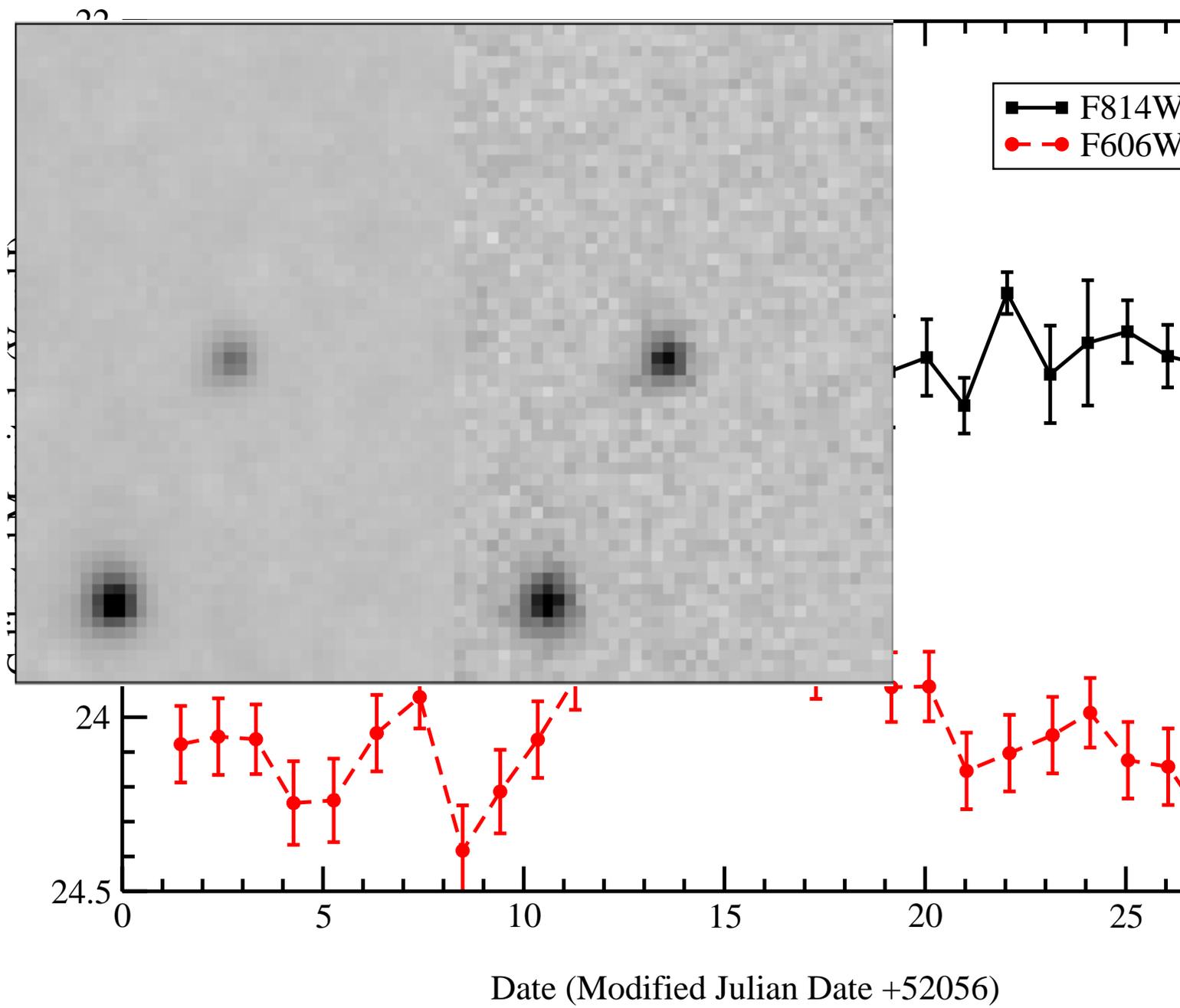
\end{document}